\documentstyle[12pt]{article}
\newcommand{\be}{\begin{equation}}
\newcommand{\ee}{\end{equation}}
\newcommand{\bea}{\begin{eqnarray}}
\newcommand{\eea}{\end{eqnarray}}
\newcommand{\bd}{\begin{displaymath}}
\newcommand{\ed}{\end{displaymath}}
\newcommand{\bi}{\begin{itemize}}
\newcommand{\ei}{\end{itemize}}
\newcommand{\bc}{\begin{center}}
\newcommand{\ec}{\end{center}}
\newcommand{\bfl}{\begin{flushleft}}
\newcommand{\efl}{\end{flushleft}}
\newcommand{\bfr}{\begin{flushright}}
\newcommand{\efr}{\end{flushright}}
\newcommand{\f}{\frac}

\def\6{\partial} \def\a{\alpha} 
\def\g{\gamma} \def\d{\delta}  \def\e{\epsilon}
  
  \def\l{\lambda}
\def\m{\mu} \def\n{\nu} \def\x{\xi} \def\p{\pi}
\def\r{\rho}  \def\t{\tau}

\def\o{\omega} \def\G{\Gamma} 
  \def\S{\Sigma}
  \def\O{\Omega}
\def\P{\Pi}

\def\={\!\!\!&=&\!\!\!}
\def\+{\!\!\!&&\!\!\!+~}
\def\-{\!\!\!&&\!\!\!-~}

\newcommand{\DD}{{\cal D}}

\newcommand{\FF}{{\cal F}}
\newcommand{\GG}{{\cal G}}

\newcommand{\OO}{{\cal O}}

\newcommand{\SS}{{\cal S}}

\begin{document}

\title{On the Dirac Eigenvalues as Observables of the on-shell $N=2$ $D=4$ Euclidean Supergravity}
\author{Ion V. Vancea\thanks{email:ionvancea@ufrrj.br}}
\date{12 March 2008}
\maketitle

\pagestyle{empty}

\begin{center}
\emph{Departamento de F\'{\i}sica, Universidade Federal Rural do Rio de
Janeiro (UFRRJ),\newline Cx. Postal 23851, 23890-000 Serop\'{e}dica - RJ,
Brasil}
\end{center}

\abstract{We generalize previous works on the Dirac eigenvalues as dynamical variables of the Euclidean gravity and $N=1$ $D=4$ supergravity to on-shell $N=2$ $D=4$ Euclidean supergravity. The covariant phase space of the theory is defined as as the space of the solutions of the equations of motion modulo the on-shell gauge transformations. In this space we define the Poisson brackets and compute their value for the Dirac eigenvalues.}

\newpage

\section{Introduction}

One of the most important and longstanding problems in the general relativity and its supersymmetric extensions is 
finding a complete set of observables of these theories. In particular, such objects should contain information about spacetime manifold and have vanishing Poisson brackets with the constraints of the general relativity. Therefore, they might be useful for quantizing the gravity and the supergravity by employing the canonical methods \cite{pb,cji}. 
Several years ago, it was shown in the framework of the noncommutative geometry that there is a relationship between 
the algebra generated by the Dirac operator and the smooth functions on a manifold and the geometry of the latter \cite{ac}. On the other hand, it is known from the spectral geometry that the eigenvalues of certain differential operators contain information on the geometry of the manifold on which they are defined \cite{pg}. These observations were applied to the general relativity in \cite{rpcr} where it was proposed that the eigenvalues of the Weyl operator played the sought for role in the Euclidean general relativity in the Ashtekar variables. However, there are two major problems in making this idea work: the difficulty of calculating the spectrum of the Weyl operator and the uniqueness of the gauge variations of theory in terms of the Ashtekar variables. Since the theory of the Dirac operator is better known, it has been put forward in \cite{lr,lr1,gl0} that the eigenvalues of the Dirac operator be considered as 
dynamical variables of the gravitational field in the Euclidean general relativity. The theory is restricted to the compact Euclidean manifolds since the Dirac operator should be hermitean and elliptic. Also, it is precisely in the Euclidean case that one can make the connection with the noncommutative geometry. In particular, the Hilbert-Einstein action of the general relativity can be written solely in terms of the eigenvalues of the Dirac operator as a Dixmier
trace \cite{lr} (see for noncommutative geometry, e. g. \cite{ac,gl} and for connection with spaces with torsion
\cite{sv1,sv2,sv3,sv4,sv5,sv6}.) 

In a previous work, the above construction was extended to the case in which, beside the diffeomorphisms and $SO(4)$ rotations, local $N=1$ supersymmetry transformations have been taken into account \cite{ivv1}. The relations resulting 
from the variation of the Dirac eigenvalues were considered {\em a priori} as constraints on the geometry of 
the Euclidean manifolds and some consequences of this hypothesis were analyzed in \cite{ivv2,ivv3,ivv4,ivv5} 
(see also the review in \cite{ivv6}). In this article, we are going to consider the on-shell $N=2$ $D=4$ Euclidean supergravity. The reasons for which one would like to understand better the Euclidean supergravity come mainly from the AdS/CFT correspondence in string theory (see \cite{rep} and the references therein) and the study of instantons in the field theories coupled with gravity. 

In the Euclidean theory, the complex Dirac spinor can be written as two Majorana spinors defined by a new Majorana condition and the bosonic fields are real \cite{pnaw1,pnaw2,bvpn}. This allows us to write local supersymmetry 
transformations consistent with $SO(4)$ as in \cite{utpn} \footnote{For different approaches to the Euclidean supersymmetry see the review in \cite{pnaw2} and the references therein.}. Our main aim is to extend the formulation of the $N=1$ $D=4$ Euclidean supergravity in which the dyamical variables are given by the Dirac eigenvalues $\l_n$'s to the on-shell $N=2$ $D=4$ Euclidean supergravity. To this end, one has to make use of the concept of the covariant phase space of supergravity. The key objects defined on it are the covariant Poisson brackets. We define the Poisson brackets for the $N=2$ $D=4$ Euclidean supergravity and derive their expression for the Dirac eigenvalues.

The plan of this paper is as follows. In the Section 2 we discuss the on-shell $N=2$ $D=4$ Euclidean supergravity in terms of
the Dirac eigenvalues. In Section 3 we define the Poisson brackets in the covariant phase space defined as the sapce of the solutions of the equations of motion modulo the local gague transformations. 
The Poisson brackets of the Dirac eigenvalues are computed in the Section 4. The last section is devoted to conclussions. The units in this paper are such that $32\p G = 1$.

\section{Dirac Eigenvalues as Dynamical Variables of $N=2$ $D=4$ Euclidean Supergravity}

Consider the $N=2$ $D=4$ Euclidean supergravity on a compact spacetime manifold $M$ without boundary. The gravitational
supermultiplet has the following field content: the real vielbein $e^{a}_{\m}$, the real vector field $A_{\m}$ and the 
gravitino $\psi^{\a}_{\m}$. The Euclidean spinors are obtained from the complex Majorana spinor $\psi^{M \a}_{\m}$ 
in the Minkowski spacetime by performing the following Wick rotation 
\cite{pnaw1,pnaw2}
\be
\psi^M \rightarrow O\psi ~~,~~ \psi^{M\dagger} \rightarrow \psi^{\dagger}O ~~,~~O=e^{\g^4 \g^5 \pi/4}.
\label{wickrotation}
\ee 
The matrix $O$ defined above acts on the Dirac matrices $\g^M$ in the Minkowski spacetime and transforms them 
into matrices in the Euclidean spacetime according to the rule
\be
\g^{M a} \rightarrow O^{-1}\g^{a}O. 
\label{diracmatrices}
\ee
The action of the $N=2$ $D=4$ Euclidean supergravity takes the following form \cite{utpn}
\begin{eqnarray}
S &=&\int d^{4}x\left[ -\frac{e}{2\kappa ^{2}}R\left( e,\omega \right)
-\varepsilon ^{\mu \nu \rho \sigma }\psi _{\mu }^{\dag }\gamma _{\nu }\nabla%
_{\rho }\psi _{\sigma }+\frac{1}{4}eF^{2}-\frac{6g^{2}}{\kappa ^{4}}%
e\right.  \\
&&\left. -\frac{i\kappa }{2\sqrt{2}}\psi _{\mu }^{\dag }\gamma _{5}\left(
e\left( F^{\mu \nu }+{\em F}^{\mu \nu }\right) +\gamma _{5}\left( \widetilde{%
F}^{\mu \nu }+\widetilde{{\em F}}^{\mu \nu }\right) +\frac{4ig}{\kappa ^{2}}%
e\gamma ^{\mu \nu }\right) \psi _{\nu }\right] ,  
\label{action}
\end{eqnarray}
where 
\begin{eqnarray}
{\em F}^{\mu \nu } &=&F^{\mu \nu }-\frac{i\kappa }{\sqrt{2}}\left( \psi
_{\mu }^{\dag }\gamma _{5}\psi _{\nu }-\psi _{\nu }^{\dag }\gamma _{5}\psi
_{\mu }\right) ,\quad \widetilde{{\em F}}^{\mu \nu }=\frac{1}{2}\varepsilon
^{\mu \nu \rho \sigma }\widetilde{{\em F}}_{\rho \sigma },  \label{F-field}
\\
\omega _{\mu ab} &=&\omega _{\mu ab}(e)-\frac{\kappa ^{2}}{4}\left( \psi
_{a}^{\dag }\gamma _{5}\gamma _{\mu }\psi _{b}+\psi _{a}^{\dag }\gamma
_{5}\gamma _{b}\psi _{\mu }-\psi _{a}^{\dag }\gamma _{5}\gamma _{b}\psi
_{a}-\left( a\longleftrightarrow b\right) \right) ,  \label{spin-conn}
\end{eqnarray}%
are the supercovariant field strength, its Hodge dual and the spin connection, respectively.
The phase space of the theory is a (super)covariant notion and it is defined as the space of the solutions of the 
equations of motion $\left( e, A, \psi \right)$ modulo the gauge transformations. Thus, it is sufficient to consider only the on-shell supergravity transformations \cite{utpn}
\bea
\delta e_{\mu }{}^{a}& =&\xi ^{\nu }\partial _{\nu }e_{\mu }{}^{a}+\partial
_{\mu }\xi ^{\nu }e_{\nu }{}^{a}-\xi _{b}{}^{a}e_{\mu }{}^{b}+\frac{1}{2}%
\,(\psi _{\mu }^{\dag }\gamma _{5}\gamma ^{a}\epsilon -\epsilon ^{\dag
}\gamma _{5}\gamma ^{a}\psi _{\mu })  
\nonumber \\
\delta A_{\mu }& =&\xi ^{\nu }\partial _{\nu }A_{\mu }+\partial _{\mu }\xi
^{\nu }A_{\nu }+\partial _{\mu }\xi +\frac{\imath }{\sqrt{2}}\,(\epsilon
^{\dag }\gamma _{5}\psi _{\mu }-\psi _{\mu }^{\dag }\gamma _{5\epsilon }\xi )
\nonumber \\
\delta \psi _{\mu }& =&\xi ^{\nu }\partial _{\nu }\psi _{\mu }+\partial _{\mu
}\xi ^{\nu }\psi _{\nu }+\frac{1}{4}\xi ^{ab}\gamma _{ab}\psi _{\mu }-^{{-1}%
}\nabla _{\mu }\epsilon +\frac{\imath }{2\sqrt{2}}\,\gamma ^{\nu }G_{\mu
\nu }\epsilon   \nonumber \\
&-&\frac{1}{\sqrt{2}}\,g^{{-2}}\gamma _{\mu }\epsilon +\imath g\xi
\,\psi _{\mu }  
\nonumber \\
\delta \psi _{\mu }^{\dag }& =&\xi ^{\nu }\partial _{\nu }\psi _{\mu }^{\dag
}+\partial _{\mu }\xi ^{\nu }\psi _{\nu }^{\dag }-\frac{1}{4}\xi ^{ab}\psi
_{\mu }^{\dag }\gamma _{ab}+_{\mu }^{{-1}}\nabla _{\mu }\epsilon ^{\dag }+%
\frac{\imath }{2\sqrt{2}}\,\epsilon ^{\dag }G_{\mu \nu }\gamma ^{\nu } 
\nonumber \\
& +& \frac{1}{\sqrt{2}}\,g^{{-2}}\epsilon ^{\dag }\gamma _{\mu
}-\imath g\xi \,\psi _{\mu }^{\dag }\ .  
\label{transf-action}
\eea%
Here, the parameters of the general coordinate transformations, local $SO(4)$ rotations, $U(1)$ gauge transformations and $N=2$ Euclidean supersymmetry transformations are denoted by $\xi ^{\mu },\xi ^{ab},\xi $ and $\epsilon
,\epsilon ^{\dagger }$, respectively.

The observables are functions on the phase space. Let us denote by $\FF$ the space of smooth supergravity multiplets $(e,A,\psi)$ which are composed by smooth tetrads, smooth connections on the $U(1)$ bundle over $M$ and smooth sections from the spin bundle $\S(M)$. The space of solutions of the equations of motion of the Euclidean supergravity is denoted by $\SS$ and the space of the gauge orbits of the transformations (\ref{transf-action}) in $\FF$ is denoted by $\OO$. Then the covariant phase space is $\GG$.

As was noted in \cite{lr,ivv1} a set of classical observables of the Euclidean gravity and $N=1$ Euclidean supergravity on compact manifolds is given by the eigenvalues $\{ \l_n \}$'s of the Dirac operator which define a discrete family of real-valued functions on the space of vielbeins. The functions $\l_n(e, A, \psi)$ are invariant under the gauge transformations for each $n$. Let us now apply the same idea to the $N=2$ $D=4$ Euclidean supergravity with the on-shell transformations given by the relations (\ref{transf-action}). The Dirac operator has the following form
\be
\DD = i \g^{a}e^{\m}_{a}\nabla_{\m}.
\label{diracoperator}
\ee
and it is given by the sum between the Dirac operator in the non-supersymmetric theory and two terms depending on the potential $A$ and the gravitiono $\psi$, respectively. The gauge- and Lorentz-covariant derivative of the $N=2$ $D=4$ Euclidean are obtained from the corresponding covariant derivative of the $N=2$ supergravity in the Minkowski background by a Wick rotation \cite{pnaw1,pnaw2}
\begin{equation}
\nabla_{\mu} \psi_{\nu} = \left( D_{\mu} - igA_{\mu}\right)\psi_{\nu},
\label{covariantderivative}
\end{equation}
where the operator $D_{\mu}$ has the following action on the vielbein and
the gravitino fields 
\begin{eqnarray}
D_{\mu}e^{a}_{\mu}&=& \left( \partial_{\mu} - \omega_{\mu b}^{a}
\right)e_{\nu}^{b} ,  
\label{cdvielbein} \\
D_{\mu}\psi_{\mu} &=& \left( \partial_{\mu} + \frac{1}{4}\omega_{\mu}^{ab}%
\gamma_{ab}\right)\psi_{\nu}.
\label{cdpsi}
\end{eqnarray}
If ones fixes the $A$ and $\psi$ fields, the symbol of the Dirac operator in the presence of supersymmetry is a local isomorphism
\be
L_0 : U \rightarrow Hom(S(M),S(M))~~,~~S(M)=\G (M)\times \S (M),
\label{symbdir}
\ee
where $U \in M$ is an open subset, $\G (M)$ is the $U(1)$ bundle and $\S(M)$ is the spin bundle over $M$, respectively
\cite{as}. Standard computations show that the Dirac operator is an elliptic operator on $M$ and since $M$ is compact, the operator $\DD$ has a discrete spectrum of eigenvalues 
\be
\DD \chi_n = \l_n \chi_n~~,~~n=0, 1, 2, \ldots .
\label{diraceigenvalues}
\ee
The spinor bundle $\S(M)$ can be endowed with the following scalar product by which it becomes an Hilbert space
\be
\left< \psi | \chi \right> = \int d^{4}x ~\sqrt{g(x)}~ \psi^{*} (x) \chi (x).
\label{scalarproduct}
\ee
Here, ${}^{*}$ denotes the complex conjugation. The $\l_n $'s define a discrete family of functions on $\FF$ which is a consequence of the dependence of $D$ on $(e, A, \psi )$
\bea
\l_{n} &:& \FF \longrightarrow R ~~~~,~~(e, A, \psi ) \rightarrow
\l_{n}(e, A, \psi ), \label{functs}\\
\l_{n} &:& \FF \longrightarrow R^{\infty}~~,~~(e, A, \psi ) \rightarrow
\{ \l_{n}(e, A, \psi ) \}, \label{seqs}
\eea
where the second correspondence defines the infinite family of dynamical variables.

The functions (\ref{functs}) and (\ref{seqs}) define a set of variables of the Euclidean supergravity.
In principle, they could be found by solving the spectral problem (\ref{diraceigenvalues}) on
arbitrary backgrounds which is a notorious difficult problem. Nevertheless, one can
obtain interesting information about the eigenvalues $\l_n$'s by investigating their variation with
respect to the variables $\left( e, A, \psi \right)$. In this way, one can check out explicitly the gauge
invariance of the eigenvalues of the Dirac operator and one can attempt to construct their Poisson 
brackets. Note that the eigenvalues of the Dirac operator should be invariant under the transformations (\ref{transf-action}) since $\DD$ is covariant by construction. 

\section{On-Shell Poisson Brackets of $N=2$ $D=4$ Euclidean Supergravity}

Let us study now the Poisson brackets on the covariant phase space of the on-shell Euclidean supergravity. A vector field on 
$\SS$ can be written as
\be
X = \int d^4 x \left( X^{a}_{\m}(x)(e,A,\psi) \f{\d}{\d e^{a}_{\m}(x)}
+ X^{\m}(x)(e,A,\psi)\f{\d}{\d A^{\m}(x)} + X^{I}_{\m}(x)(e,A,\psi)\f{\d}{\d\psi^{I}_{\m}(x)}
\right),
\label{vectfield}
\ee
where $X^{a}_{\m}(x),X^{\m}(x)$ and $X^{I}_{\m}(x)$ are solutions of the equations of motion of $N=2$
$D=4$ supergravity. The vector field $X$ defines a vector field $[ X ]$ on the covariant phase space $\GG$ as the 
equivalence class of (\ref{vectfield}) modulo the linearized gauge transformations. The symplectic two-form of the general relativity from \cite{abr} can be generalized to the supergravity by defining the following two-form field
\be
\O(X,Y) = \int_{N} d^3 \sigma ~n_{\r}\left( X^{a}_{\m} \stackrel{\rightleftharpoons}{\nabla_{\sigma}}Y^{b}_{\n}
\epsilon^{\sigma}_{ab\t} + X_{\m}\stackrel{\rightleftharpoons}{\nabla_{\t}} Y_{\n} +
X^{I}_{\m}\stackrel{\rightleftharpoons}{\nabla_{\t}}Y^{J}_{\n}\e_{IJ}
\right)\e^{\t \r \m \n},
\label{twoformsugra}
\ee
where $N$ is a compact topological three-dimensional surface such that $M = N \times S^1$,  $n_{\r}$ is its 
normal one-form and
\be
F \stackrel{\rightleftharpoons}{\nabla_{\sigma}} G = F \nabla_{\sigma} G - G \nabla_{\sigma}F. 
\label{stack}
\ee
Since the solutions of the Euclidean supergravity depends on the full supergravity multiplet, the covariant derivative
that from (\ref{twoformsugra}) is the one that acts on the spinors (\ref{cdpsi}). The components of $\O$ are defined in each sector by the following relations
\bea
\O^{\m \n}_{ab}(x,y) &=& \int_{N}d^3 \sigma ~n_{\r}\d(x,\sigma(x))\stackrel{\rightleftharpoons}{\nabla_{\sigma}}\d (y,x(\sigma))
\e^{\r}_{ab\t}\e^{\t\sigma \m\n}\label{compo1},\\
\O^{\m \n}(x,y) &=& \int_{N}d^3 \sigma ~n_{\r}\d(x,\sigma(x))\stackrel{\rightleftharpoons}{\nabla_{\sigma}}\d (y,x(\sigma))
\e^{\r\sigma\m\n}\label{compo2},\\
\O^{\m \n}_{IJ} (x,y)&=& \int_{N}d^3 \sigma ~n_{\r}\d(x,\sigma(x))\stackrel{\rightleftharpoons}{\nabla_{\sigma}}\d (y,x(\sigma))
\e_{IJ}\e^{\r\sigma\m\n},\label{compo3}
\eea
where $N: \sigma \rightarrow \x(\sigma)$. Like 
the symplectic two-form of gravity which can be recongnized in the first term of (\ref{twoformsugra}), $\O$ is
degenerate in the gauge directions and has an inverse in the space of the orbits. Nevertheless, once the gauge fields
and a representative of the vector field $[X]$ are fixed, then $\O$ admits an inverse $\P$. The components of $\O$ are
defined by the following relations
\bea
\int d^4 y \int d^4 z ~\P^{ab}_{\m\n}(x,y)\O^{\n\r}_{bc}(y,z)X^{c}_{\r}(z)&=&\int d^4z ~\d(x,z)\d^{\r}_{\m}\d^{a}_{c}
X^{c}_{\r}(z)\label{inverse1},\\
\int d^4 y \int d^4 z ~\P_{\m\n}(x,y)\O^{\n\r}(y,z)X_{\r}(z)&=&\int d^4z ~\d(x,z)\d^{\r}_{\m}
X_{\r}(z)\label{inverse2},\\
\int d^4 y \int d^4 z ~\P^{IJ}_{\m\n}(x,y)\O^{\n\r}_{JK}(y,z)X^{K}_{\r}(z)&=&\int d^4z ~\d(x,z)\d^{\r}_{\m}\d^{I}_{K}
X^{K}_{\r}(z)\label{inverse3},
\eea
where $\{ X^{c}_{\r},X_{\r},X^{K}_{\r} \}$ denote an arbitrary solution of the supergravity equations of motion which 
satisfies the chosen gauge. By using the relations (\ref{compo1}), (\ref{compo2}) and (\ref{compo3}) in to 
(\ref{inverse1}), (\ref{inverse2}) and (\ref{inverse3}) one obtains a set of equations that defines the components
of the inverse $\P$ in the three sectors
\bea
\int_{N}d^3\sigma ~n_{\r} \left( \P^{ab}_{\m\n}(x,x(\sigma)) \stackrel{\rightleftharpoons}{\nabla_{\sigma}} 
X^{c}_{\g}(x(\sigma)) \right) \e^{\r}_{bc\t}\e^{\t\n\sigma\g}  &=&X^{a}_{\m}(x)
\label{eqinv1},\\
\int_{N}d^3\sigma ~n_{\r} \left( \P_{\m\n}(x,x(\sigma)) \stackrel{\rightleftharpoons}{\nabla_{\sigma}} 
X_{\g}(x(\sigma)) \right) \e^{\r\n\sigma\g}  &=&X_{\m}(x)
\label{eqinv2},\\
\int_{N}d^3\sigma ~n_{\r} \left( \P^{IJ}_{\m\n}(x,x(\sigma)) \stackrel{\rightleftharpoons}{\nabla_{\sigma}} 
X^{K}_{\g}(x(\sigma)) \right) \e_{JK}\e^{\r\n\sigma\g}  &=&X^{I}_{\m}(x)
\label{eqinv3}.
\eea
The Poisson brackets of two functions $F,G$ on the space $\SS$ can be written as
\bea
\{ F, G \} &=& \int d^4 x \int d^4 y \left( \P^{ab}_{\m\n}(x,y)\f{\d F}{\d e^{a}_{\m}(x)}\f{\d G}{\d e^{b}_{\n}(y)}
+ \P_{\m\n}(x,y)\f{\d F}{\d A_{\m}(x)}\f{\d G}{\d A_{\n}(y)} \right.
\nonumber \\
&+& \left. \P^{IJ}_{\m\n}(x,y)\f{\d F}{\d \psi^{I}_{\m}(x)}
\f{\d G}{\d \psi^{J}_{\n}}\right).
\label{poisson}
\eea
Note that the relations (\ref{eqinv1}), (\ref{eqinv2}) and (\ref{eqinv3}) define the extension 
of the propagator of the linearized Einstein equations \cite{lr1} to the linearized $N=2$ $D=4$ Euclidean 
supergravity in the chosen gauge. The Poisson brackets from (\ref{poisson}) depend on the gauge, too. However, if
$F$ and $G$ are gauge invariant functions then the Poisson brackets are gauge independent.

\section{Poisson Brackets of the Dirac Eigenvalues}

The local variation of the Dirac eigenvalues is given by the perturbation theory
\be
\delta \l_n = \left \langle \chi_n \right| \delta \DD \left| \chi_n \right \rangle .
\label{variation}
\ee
The terms containing the local variation of the eigenspinors $\chi_n$ cancel each other. The variation of the 
Dirac eigenvalues under gauge transformations has the following form
\bea
\d \l_n &=& i \left\langle \chi_n \right| \g^a \d e^{\m}_a \nabla_{\m} \left| \chi_n \right \rangle
+\f{i}{4}\left\langle \chi_n \right| \g^a e^{\m}_a  \d \o_{\m}^{cd} \g_{cd}\left| \chi_n \right \rangle
+ g\left\langle \chi_n \right| \g^a e^{\m}_a \d A_{\m} \left| \chi_n \right \rangle \nonumber\\
&=&
i \left\langle \chi_n \right| \g^a \left( \xi^{\n}\6_{\n}e_{\m}^a + \6_{\m}\xi^{\n}e_{\n}^{a}
-\xi^{a}_{b}e_{\m}^{b} + \f{1}{2}k \left( \psi^{\dagger}_{\m}\g_5 \g^{a}\e -\e^{\dagger}\g_5\g^a\psi_{\m} 
\right) \right)\nabla^{\m}\left| \chi_n \right \rangle \nonumber\\
&+&\f{i}{4}\left\langle \chi_n \right| \g^a e^{\m}_a \d \o_{\m}^{cd} \g_{cd} \left| \chi_n \right \rangle \nonumber\\
&+&g\left\langle \chi_n \right| \g^a e^{\m}_a \left( \xi^{\n}\6_{\n}A_{\m} + \6_{\m}\xi^{\n}A_{\n}
+ \6_{\m}\xi + \f{i}{\sqrt{2}}\left( \e^{\dagger} \g_5 \psi_{\m} - \psi^{\dagger}_{\m}\g_5\e \right)
 \right) \left| \chi_n \right \rangle \label{varlambda},
\eea
where the variation of the spin connection is given by the following relation
\bea
\d \o_{\m}^{ab} & = & \d \o_{\m}^{ab}(e) 
-\f{k^2}{4}\left[ \left( \d e^{a}_{\n} \psi^{\n \dagger } + e^{a}_{\n} \d \psi^{\n \dagger} \right)\g_5\g_{\m}\psi^b 
+\psi^{a\dagger}\g_5\g_{\m}\left( \d e^{b}_{\n} \psi^{\n \dagger } + e^{b}_{\n} \d \psi^{\n \dagger} \right)\right]
\nonumber\\
&-&\f{k^2}{4}\left[
\left( \d e^{a}_{\n} \psi^{\n \dagger } + e^{a}_{\n} \d \psi^{\n \dagger} \right)\g_5\g^{b}\psi_{\m}
+
\psi^{a\dagger}\g_5\g^{b}\d \psi_{\m}
\right]
\nonumber \\
&+&\f{k^2}{4}\left[ \d \psi^{\dagger}_{\m}\g_5 \g^{a}\psi^{b} - \psi^{\dagger}_{\m}\g_5\g^a
\left( \d e^{b}_{\n}\psi^{\n} + e^{b}_{\n}\d \psi^{\n}\right)\right] \nonumber\\
&-& (a\leftrightarrow b).\nonumber
\eea
If the explicit form of the gauge transformations from (\ref{transf-action}) \cite{utpn} is employed, one obtains the following variation of the Dirac eigenvalues 
\bea
\d \l_n &=&\d \o_{\m}^{ab} (e) \nonumber\\
&-&\f{k^2}{4}\left[ 
\left( \xi^{\r}\6_{\r}e_{\n}^c + \6_{\n}\xi^{\r}e_{\r}^{c}
-\xi^{c}_{d}e_{\n}^{d} + \f{k}{2}\left( \psi^{\dagger}_{\n}\g_5 \g^{c}\e -\e^{\dagger}\g_5\g^c\psi_{\n} 
\right) \right)
\left( \d^{ca}\psi^{\n\dagger}\g_5\g_{\m}\psi^b + \psi^{a\dagger}\g_5\g_{\m}\psi^{\n \dagger}\d^{cb} 
\right)
\right.\nonumber\\
&+&
\left( \xi^{\r}\6_{\r}\psi^{\n \dagger} + \6^{\n}\xi^{\r}\psi^{\n \dagger} - \f{1}{4}\xi^{cd}\psi^{\n \dagger}
\g_{cd} +\f{\nabla^{\n}\xi^{\dagger}}{k} + \f{i \xi^{\dagger}G^{\n\r}\g_{\r}}{2\sqrt{2}} + 
\f{g\xi^{\dagger}\g^{\n}}{\sqrt{2}} -ig\xi\psi^{\n \dagger}\right)e^{a}_{\n}\g_5\g_{\m}\psi^{b}
\nonumber\\
&+&\left.\psi^{a\dagger}\g_5\g_{\m}e^{b}_{\n}\left( \xi^{\r}\6_{\r}\psi^{\n \dagger} + \6^{\n}\xi^{\r}\psi^{\n \dagger} - \f{1}{4}\xi^{cd}\psi^{\n \dagger}
\g_{cd} +\f{\nabla^{\n}\xi^{\dagger}}{k} + \f{i \xi^{\dagger}G^{\n\r}\g_{\r}}{2\sqrt{2}} + 
\f{g\xi^{\dagger}\g^{\n}}{\sqrt{2}} -ig\xi\psi^{\n \dagger}\right)\right]\nonumber\\
&-&\f{k^2}{4}
\left( \xi^{\r}\6_{\r}e_{\n}^a + \6_{\n}\xi^{\r}e_{\r}^{a}
-\xi^{a}_{d}e_{\n}^{d} + \f{k}{2}\left( \psi^{\dagger}_{\n}\g_5 \g^{a}\e -\e^{\dagger}\g_5\g^a\psi_{\n} 
\right) \right) \psi^{\n \dagger}\g_5\g^b \psi_{\m}\nonumber\\
&-&\f{k^2}{4}
\left( \xi^{\r}\6_{\r}\psi^{\n \dagger} + \6^{\n}\xi^{\r}\psi^{\n \dagger} - \f{1}{4}\xi^{cd}\psi^{\n \dagger}
\g_{cd} +\f{\nabla^{\n}\xi^{\dagger}}{k} + \f{i \xi^{\dagger}G^{\n\r}\g_{\r}}{2\sqrt{2}} + 
\f{g\xi^{\dagger}\g^{\n}}{\sqrt{2}} -ig\xi\psi^{\n \dagger}\right)e^{a}_{\n}\g_5\g^{b}\psi_{\m}\nonumber\\
&-&\f{k^2}{4}\psi^{a\dagger}\g_5\g^{b}
\left( \xi^{\n}\6_{\n}\psi_{\m} + \6_{\m}\xi^{\n}\psi_{\n}
+\f{\xi^{cd}\g_{cd}\psi_{\m}}{4}
- \f{\nabla_{\m}\e }{k}+ \f{i\g^{\n}G_{\m \n}\e}{2\sqrt{2}} 
-\f{g\g_{\m}\e}{k^2\sqrt{2}} +ig\xi\psi_{\m}
 \right)\nonumber\\
&+&\f{k^2}{4}
\left( \xi^{\r}\6_{\r}\psi^{\dagger}_{\m} + \6_{\m}\xi^{\r}\psi^{\dagger}_{\m} - \f{1}{4}\xi^{cd}\psi^{\dagger}_{\m}
\g_{cd} +\f{\nabla_{\m}\xi^{\dagger}}{k} + \f{i \xi^{\dagger}G^{\r}_{m}\g_{\r}}{2\sqrt{2}} + 
\f{g\xi^{\dagger}\g_{\m}}{\sqrt{2}} -ig\xi\psi^{\dagger}_{\m}\right)\g_5\g^a\psi^b\nonumber\\
&-&\f{k^2}{4}\psi^{\dagger}_{\m}\g_5\g^{a}\psi^{\n}
\left( \xi^{\r}\6_{\r}e_{\n}^b + \6_{\n}\xi^{\r}e_{\r}^{b}
-\xi^{b}_{d}e_{\n}^{d} + \f{k}{2}\left( \psi^{\dagger}_{\n}\g_5 \g^{b}\e -\e^{\dagger}\g_5\g^b\psi_{\n} 
\right) \right)\nonumber\\
&-&\f{k^2}{4}\psi^{\dagger}_{\m}\g_5\g^a e^{b}_{\r}
\left( \xi^{\n}\6_{\n}\psi^{\r} + \6^{\r}\xi^{\n}\psi_{\n}
+\f{\xi^{cd}\g_{cd}\psi^{\r}}{4}
- \f{\nabla^{\r}\e }{k}+ \f{i\g^{\n}G_{\n}^{\r}\e}{2\sqrt{2}} 
-\f{g\g^{\r}\e}{k^2\sqrt{2}} +ig\xi\psi^{\r}\right) \nonumber\\
&-&(a \leftrightarrow b). \label{vardiraceigen}
\eea
The above relation vanishes identically due to the gauge invariace of the theory but it can be used to check the
explicit invariance of the Dirac eigenvalues. However, $\l_n$'s are not invariant under {\em arbitrary} variations
of the supergravity multiplet. 

Consider the linearized arbitrary transformations of the fields $(e, A, \psi)$ with the real parameter $v$
\bea
e^{a}_{v\m} & = & e^{a}_{\m} + v\d e^{a}_{\m}, \label{lin1}\\
A_{v\m} & = & A_{\m} + v\d A_{\m}, \label{lin2}\\
\psi_{v\m} & = & \psi_{\m} + v\d \psi_{\m}. \label{lin3}
\eea 
The variation of the Dirac eigenvalues with respect to $v$ can be computed from the following relation
\bea
\left. \f{d \l_n(e, A, \psi)}{d v} \right|_{v=0} &=& \int d^4 x \left[ \f{\d \l_n (e, A, \psi)}{\d e^{a}_{\m}}\d e^{a}_{\m}+
\f{\d \l_n (e, A, \psi)}{\d A_{\m}}\d A_{\m} +  \f{\d \l_n (e, A, \psi)}{\d \psi_{\m}}\d \psi_{\m} \right]
\nonumber\\
&=& \langle \chi_n | \left.\f{d}{d v}\right|_{v=0} \DD | \chi_n \rangle.
\label{varlambdav}
\eea
Using the following notations
\bea
T^{\m}_{na} &=& \f{\d}{\d e^{a}_{\m}}S_n ~,~ U^{\m}_{n} = \f{\d}{\d A_{\m}}S_n ~,~
V^{I \m}_{n} = \f{\d}{\d \psi^{\m}_{I}}S_n\label{tensorsn},\\
S_n &=& \int d^4x \sqrt{\mbox{det}(g_{\m\n}})(\chi_n^* \DD \chi_n - \chi_n^* \chi_n).
\label{actionn}
\eea
one obtains the equations that give the variation of the eigenvalues $\l_n$'s under the general linear variation of the 
supergravity multiplet
\be
\f{d \l_n(e, A, \psi)}{\d e^a_{\m}} = T^{\m}_{na}~,~
\f{d \l_n(e, A, \psi)}{\d A_{\m}} = U^{\m}_{n}~,~
\f{d \l_n(e, A, \psi)}{\d \psi^{I}_{\m}} = V^{I \m}_{n}.
\label{eqvareigenval}
\ee
The interpretation of the equations (\ref{eqvareigenval}) is that they are the Jacobian matrix of the map 
(\ref{functs}). As in the  case of general relativity, the above results allow one to study the map (\ref{functs})
in the space of supermultiplets by studying the quantities defined in (\ref{tensorsn}). In principle, the quantities 
given by the above set of relations can be computed for any given supergravity background. With their help one
can compute the Poisson brackets of the Dirac eigenvalues
\bea
\{\l_n, \l_m \} &=& \int d^4x \int d^4y ~\left( T^{\m}_{na}(x) \P^{ab}_{\m\n}(x,y)T^{\n}_{mb}(y) 
+ U^{\m}_{n}(x) \P_{\m\n}(x,y)U^{\n}_{m}(y)\right.
\nonumber\\
&+& \left. V^{I \m}_{n}(x) \P^{IJ}_{\m\n}(x,y)V^{J \n}_{m}(y)
\right).
\label{poissondireig}
\eea
The relation (\ref{poissondireig}) represents the gauge independent Poisson brackets of two eigenvalues of the 
Dirac operator in
terms of the objects computed in (\ref{eqvareigenval}) and the propagator of the linearized equations of the 
$N=2$ $D=4$ on-shell supergravity. 

Like in the case of the gravity, one can express the symplectic form in the covariant phase space $\GG$ in terms of
the differentials of the Dirac eigenvalues
\be
\O = \sum_{n,m} ~\O_{mn} ~d \l_n \wedge d \l_m,
\label{direcsymplop}
\ee
where the coefficients $\O_{mn}$ should satisfy simultaneously the following set of relations
\bea
\sum_{m,n}~\O_{mn}T^{\m}_{na}(x)T^{\n}_{mb}(y) &=& \O^{\m\n}_{ab}(x,y)
\label{defdircoef1},\\
\sum_{m,n}~\O_{mn}U^{\m}_{n}(x)U^{\n}_{m}(y) &=& \O^{\m\n}(x,y)
\label{defdircoef2},\\
\sum_{m,n}~\O_{mn}V^{I \m}_{n}(x)V^{J \n}_{m}(y) &=& \O^{IJ}_{\m\n}(x,y)
\label{defdircoef3}.
\eea
The relations (\ref{defdircoef1}), (\ref{defdircoef2}) and (\ref{defdircoef3}) hold only if the maps (\ref{functs})
and (\ref{seqs}) are invertible on the phase space. 

\section{Conclussions}

In this paperwe, we have shown that the Dirac eigenvalues form a set of dynamical variables of the on-shell $N=2$ $D=4$ Euclidean supergravity and that the notions of covariant phase space and the Poisson brackets can be extended to this system. In particular, the Poisson brackets can be computed for the Dirac eigenvalues by calculating the variation of
$\l_n$'s under arbitrary transformations. The result is given in the relation (\ref{poissondireig}) and it is
gauge invariant. The coefficients of the symplectic form $\O_{mn}$ can be expressed in terms of the coefficients on the phase space if the maps (\ref{functs}) and (\ref{seqs}) are invertible. The similar question in general relativity
was addressed in \cite{ivv5} and the answer was that there is a certain indeterminacy in finding the solutions of
the equations that define the relation between the coefficients of the symplectic form in the two representations.
It would be interesting to see if the supersymmetry and $U(1)$ symmetry can lift this indeterminacy. 
As in the case of the local $N=1$ supersymmetry analysed in previous papers, the on-shell transformations define a set of maps from the space of smooth supermultiplets. These maps lead to a set of equations involving geometrical objects associated with the spacetime manifold which are interpreted as constraints on spacetime. When the maps are identified with the Dirac eigenvalues, the equations should vanish identically since $\lambda^n$'s are gauge invariant. Nevertheless, the constraints are still left behind and they relate the geometrical objects in spacetime.
Another way to study te relationship between the Dirac eigenvalues and the geometry of the spacetime manifold is by using arbitrary transformations. Then a constraint is generated for any transformation for which $\d \l_n$'s vanish
\be
T^{\m}_{na}\d e^a_{\m} + U^{\m}_{n}\d A_{\m} + V^{I \m}_{n}\d \psi^{I}_{\m} = 0.
\label{constraint}
\ee
However, it is not know, in general, whether such transformations exist and if they are compatible with the local supersymmetry.

{\bf Acknowledgments} I would like to thank to J. A. Helayel-Neto for useful discussions and to A. M. O. de Almeida and S. A. Dias for hospitality at LAFEX-CBPF where part of this paper was done. A preliminary version of this work was partially supported by FAPESP Grant 02/05327-3.

\end{document}